\documentclass[12pt]{article}
\usepackage[english,german,french,polish]{babel}
\usepackage[T1]{fontenc}
\usepackage{amsfonts}

\begin{document}

\selectlanguage{english}

\baselineskip 0.76cm
\topmargin -0.6in
\oddsidemargin -0.1in

\let\ni=\noindent

\renewcommand{\thefootnote}{\fnsymbol{footnote}}

\newcommand{\SM}{Standard Model }

\pagestyle {plain}

\setcounter{page}{1}

\pagestyle{empty}

~~~

\begin{flushright}
IFT-- 05/4
\end{flushright}

\vspace{0.4cm}

{\large\centerline{\bf Lepton generation-weighting factors and neutrino mass formula}} 

\vspace{0.4cm}

{\centerline {\sc Wojciech Kr\'{o}likowski}}

\vspace{0.3cm}

{\centerline {\it Institute of Theoretical Physics, Warsaw University}}

{\centerline {\it Ho\.{z}a 69,~~PL--00--681 Warszawa, ~Poland}}

\vspace{0.3cm}

{\centerline{\bf Abstract}}

\vspace{0.2cm}

A candidate for the simple empirical neutrino mass formula is found, predicting the mass proportion $m_1 : m_2 : m_3 = 0  : 4  : 24$ and so, the mass ratio $\Delta m^2_{32}/\Delta m^2_{21} = 35$ not 
inconsistent with its experimental estimate. It involves only one free parameter and three 
generation-weighting factors suggested by the successful mass formula found previously for 
charged leptons (the simplest neutrino mass formula would predict $m_1 : m_2 : m_3 = 1  : 4  : 24$ 
and thus, $\Delta m^2_{32}/\Delta m^2_{21} \simeq 37$). A more involved variation of this equation follows from a special seesaw neutrino model with specifically "conspiring" ~Dirac and Majorana neutrino mass matrices. In this variation $m_1 : m_2 : m_3 \simeq \varepsilon^{(\nu)}  : 4  : 24$, where 
$O(\varepsilon^{(\nu)}) = 10^{-3}$. 

\vspace{0.6cm}

\ni PACS numbers: 12.15.Ff , 14.60.Pq , 12.15.Hh .

\vspace{0.6cm}

\ni March 2005  

\vfill\eject

~~~
\pagestyle {plain}

\setcounter{page}{1}

\vspace{0.2cm}

Some time ago we found an efficient empirical mass formula for charged leptons $e_i = e^-, \mu^-, \tau^-$ [1]. This formula reads

\begin{equation}
m_{e_i} = \mu^{(e)} \rho_i  \left(N^2_i + \frac{\varepsilon^{(e)} - 1}{N^2_i} \right) \,,
\end{equation}

\ni where 

\begin{equation}
N_i = 1,3,5 \;,
\end{equation}

\ni and

\begin{equation}
\rho_i = \frac{1}{29} \,,\,\frac{4}{29} \,,\,\frac{24}{29} 
\end{equation}

\ni ($\sum_i \rho_i = 1$). Here, $\mu^{(e)} > 0$ and $\varepsilon^{(e)} > 0$ are constants. In fact, with the experimental values $m_e = 0.510999$ MeV and $m_\mu = 105.658$ MeV as an input, the formula (1), rewritten explicitly as

\begin{equation}
m_e = \frac{\mu^{(e)}}{29} \varepsilon^{(e)} \;,\;  m_\mu = \frac{\mu^{(e)}}{29} \frac{4}{9} (80 +  \varepsilon^{(e)}) \;,\;  m_\tau = \frac{\mu^{(e)}}{29} \frac{24}{25} (624 + \varepsilon^{(e)})\;,
\end{equation}

\ni leads to {\it the prediction}

\begin{equation}
m_\tau = \frac{6}{125}(351 m_\mu - 136 m_e) = 1776.80 \;{\rm MeV}
\end{equation}

\ni and also determines both constants

\begin{equation}
\mu^{(e)} = \frac{29(9m_\mu - 4m_e)}{320} = 85.9924 \;{\rm MeV} \;,\; \varepsilon^{(e)} = \frac{320 m_e}{9m_\mu - 4m_e} = 0.172329 \,.
\end{equation}

\ni The prediction (5) is really close to the experimental value $ m^{\rm exp}_\tau = 1776.99^{+0.29}_{-0.26}$ MeV [2]. 

Though the formula (1) has essentially the empirical character, there exists a speculative background for it based on a K\"{a}hler-like extension of Dirac equation that the interested reader may find in Ref. [1]. In particular, the numbers $N_i$ and $\rho_i \; (i=1,2,3)$ given in Eqs. (2) and (3) are interpreted there. Let us only mention that $N_i - 1 = 0,2,4$ is the number of {\it additional} bispinor indices appearing in the extended Dirac equation and obeying Fermi statistics that enforces their antisymmetrization and so, restricts to zero the related additional spin. This Fermi statistics is also the reason, why there are precisely {\it three} \SM fermion generations {\it i.e.}, $N_i - 1 = 0,2,4$, since any additional bispinor index can assume {\it four} values, what implies that $N_i - 1 \leq 4$ (after the antisymmetrization of additional bispinor indices). Thus, an analogue of Pauli principle works (intrinsically), restricting the number of additional bispinor indices to $\leq 4$ and so, resulting into three and only three generations of leptons and quarks (all with spin 1/2). The generation-weighting factors $\rho_i$ multiplied by 29, $ 29\rho_i = 1,4,24$ ($\sum_i \rho_i = 1$), tell us, how many times the lepton or quark wave functions of three generations are realized (up to the factor $\pm1$) by the extended Dirac equation.

Now, it is tempting to seek in the same framework an efficient empirical mass formula for mass neutrinos $\nu_i = \nu_1,\nu_2,\nu_3 $ (being the mass states of the flavor neutrinos $\nu_\alpha = \nu_e,\nu_\mu,\nu_\tau $).

As is well known, the mass neutrinos display a less hierarchical spectrum than the charged leptons. In fact, neutrino oscillation experiments give actually the following estimates [3] for $\Delta m^2_{ji} \equiv m^2_{\nu_j} - m^2_{\nu_i}$ : the ranges

\begin{equation}
7.2 < \Delta m^2_{21}/(10^{-5}\;{\rm eV}^2) < 9.1\;,\; 1.9 < \Delta m^2_{32}/(10^{-3}\;{\rm eV}^2) < 3.0 
\end{equation}

\ni and the best fits

\begin{equation} 
\Delta m^2_{21} \sim 8.1\times 10^{-5}\;{\rm eV}^2 \;,\; \Delta m^2_{32} \sim 2.4 \times10^{-3} \;{\rm eV}^2. 
\end{equation}

\ni For the values (8) $\Delta m^2_{32}/\Delta m^2_{21} \sim 30$. Notice that 1.9/0.091 $\sim 21$ and 3.0/0.072 $\sim 42$ and so, the experimental limits are $21 < \Delta m^2_{32}/\Delta m^2_{21} < 42$.

Thus, let us tentatively try for the neutrino mass formula the simplest conjecture

\begin{equation}
m_{\nu_i} = \mu^{(\nu)} \rho_i \,,
\end{equation}

\ni where the generation-weighting factors $\rho_i $ as given in Eq. (3) still appear, while the numbers $N_i$ numerating the generations and defined in Eq. (2) are absent. Here, $\mu^{(\nu)} > 0$ is a constant.

The tentative mass formula (9), rewritten as

\begin{equation}
m_{\nu_1} = \frac{1}{29}\mu^{(\nu)}  \,,\, m_{\nu_2} = \frac{4}{29}\mu^{(\nu)}  \,,\, m_{\nu_3} = \frac{24}{29}\mu^{(\nu)}  \,,
\end{equation}

\ni implies that

\begin{equation}
m_{\nu_1} : m_{\nu_2} : m_{\nu_3}  = 1 : 4 : 24
\end{equation}

\ni and

\begin{equation}
\mu^{(\nu)} = m_{\nu_1} + m_{\nu_2} + m_{\nu_3} = 29m_{\nu_1} = \frac{29}{4} m_{\nu_2}  = \frac{29}{24} m_{\nu_3} \,. 
\end{equation}

\ni From Eq. (11)

\begin{equation}
\Delta m^2_{32}/\Delta m^2_{21} = \frac{112}{3} = 37.3333 \,.
\end{equation}

\ni Thus, using the experimental range (7) of $\Delta m^2_{21}$ and its experimental best fit (8) as an input, we get the following {\it prediction}: the range

\begin{equation}
2.7 < \Delta m^2_{32}/(10^{-3}\;{\rm eV}^2) < 3.4 
\end{equation}

\ni and the best fit :

\begin{equation}
\Delta m^2_{32} \sim 3.0 \times 10^{-3}\;{\rm eV}^2 \,.
\end{equation}

The predicted range (14) of  $ \Delta m^2_{32}$ is not inconsistent with its experimental range (7), but its predicted best fit (15) appears too large in comparison with the experimental best fit (8) (though the predicted ratio (13) remains within its experimental limits $21 < \Delta m^2_{32}/\Delta m^2_{21} < 42$). Note that making use of the best fit (15) for $ \Delta m^2_{32} $, we would {\it predict} from Eq. (11)

\begin{equation}
m_{\nu_1} \sim 2.3\times 10^{-3}\;{\rm eV}\,,\, m_{\nu_2} \sim 9.3\times 10^{-3}\;{\rm eV}\,,\, m_{\nu_3} \sim 5.6\times 10^{-2}\;{\rm eV}
\end{equation}

\ni and determine from Eq. (12)

\begin{equation}
\mu^{(\nu)} \sim 6.7\times 10^{-3}\;{\rm eV} \,.
\end{equation}

\ni Here, the only input is the experimental estimate (8) of $ \Delta m^2_{21}$.

We may argue that the tentative mass formula (9) requires a correction for the smallest neutrino mass $m_{\nu_1}$, if the neutrino masses are related ({\it grosso modo}) to the additional bispinor indices in the general Dirac equation applied to the neutrino triplet. Then, for the $\nu_1$ neutrino -- that does not involve additional indices -- we ought to expect $ m_{\nu_1} = 0$ (at least approximately). This conjecture may lead to the correction factor $1 - \delta_{i1}$ in the mass equation (9). In consequence, the corrected neutrino mass formula may read

\begin{equation}
m_{\nu_i} = \mu^{(\nu)} \rho_i (1 - \delta_{i1}) \,.
\end{equation}

This mass formula, rewritten as

\begin{equation}
m_{\nu_1} =0\,,\, m_{\nu_2} = \frac{4}{29} \mu^{(\nu)}\,,\, m_{\nu_3} = \frac{24}{29} \mu^{(\nu)}\,,
\end{equation}

\ni gives

\begin{equation}
m_{\nu_1} : m_{\nu_2} : m_{\nu_3} = 0 : 4 : 24
\end{equation}

\ni and

\begin{equation}
\mu^{(\nu)} = \frac{29}{28} (m_{\nu_2} + m_{\nu_3}) = \frac{29}{4} m_{\nu_2} = \frac{29}{24} m_{\nu_3}\;.
\end{equation}

\ni From Eq. (20)

\begin{equation}
\Delta m^2_{32}/\Delta m^2_{21} = 35 \;.
\end{equation}

\ni Hence, making use of the experimental range (7) of $\Delta m^2_{21}$ and its experimental best fit (8) as an input, we obtain the following {\it prediction} : the range

\begin{equation}
2.5 < \Delta m^2_{32}/ (10^{-3}\;{\rm eV}^2)< 3.2
\end{equation}

\ni and the best fit

\begin{equation}
\Delta m^2_{32} \sim 2.8 \times 10^{-3}\;{\rm eV}^2 \;.
\end{equation}

The predicted range (23) of $\Delta m^2_{32}$ is a little closer to its experimental range (7) than the previous range (14) (both being not inconsistent with (7)). Also the predicted best fit (24) is a bit closer to its actual experimental best fit (8) than the previous best fit (15) (both being too large, though the predicted ratios (13) and (22) remain within their actual experimental limits $21 < \Delta m^2_{32}/\Delta m^2_{21} <42$). Note that using the best fit (24) for $\Delta m^2_{32}$, we would {\it predict} from Eq. (20)

\begin{equation}
m_{\nu_1} \sim 0\,,\, m_{\nu_2} \sim 9.0 \times 10^{-3}\;{\rm eV} \,,\, m_{\nu_3} \sim 5.4 \times 10^{-2}\;{\rm eV} 
\end{equation}

\ni and determine from Eq. (21)

\begin{equation}
\mu^{(\nu)} \sim 6.5 \times 10^{-2}\;{\rm eV} \;.
\end{equation}

\ni The experimental best fit (8) for $\Delta m^2_{21}$ is the only input here.

Naturally, the actual experimental best fit $\Delta m^2_{21} \sim 8.1\times 10^{-5}\;{\rm eV}^2$ (giving $\Delta m^2_{32} \sim 3.0\times 10^{-3}\;{\rm eV}^2$ through Eq. (22)) may change in the course of further experiments. For instance, if (drastically) it turned out as small as $\Delta m^2_{21} \sim (6.9 - 7.2)\times 10^{-5}\;{\rm eV}^2$, we would predict from Eq. (22) that $\Delta m^2_{32} \sim (2.4 - 2.5)\times 10^{-3}\;{\rm eV}^2$. Then, from Eq. (20)

\begin{equation}
m_{\nu_1} \sim 0 \;,\; m_{\nu_2} \sim (8.3 - 8.5) \times 10^{-3}\;{\rm eV} \;,\; m_{\nu_3} \sim (5.0 - 5.1) \times 10^{-2}\;{\rm eV} 
\end{equation}

\ni and from Eq. (21)

\begin{equation}
\mu^{(\nu)} \sim (6.0 - 6.1) \times 10^{-2}\;{\rm eV}\;.
\end{equation}

\ni Similarly, the actual experimental best fit $\Delta m^2_{32} \sim 2.4\times 10^{-3}\;{\rm eV}^2$ (giving $\Delta m^2_{21} \sim 6.9 \times 10^{-5}\;{\rm eV}^2$ by means of Eq. (22)) may change. For example, if (drastically) it appeared as large as $\Delta m^2_{32} \sim (2.8 - 2.9)\times 10^{-3}\;{\rm eV}^2$, we would predict from Eq. (22) that $\Delta m^2_{21} \sim (8.0 - 8.3)\times 10^{-5}\;{\rm eV}^2$. Then, from Eq. (20)

\begin{equation}
m_{\nu_1} \sim 0 \;,\; m_{\nu_2} \sim (9.0 - 9.1) \times 10^{-3}\;{\rm eV} \;,\; m_{\nu_3} \sim (5.4 - 5.5) \times 10^{-2}\;{\rm eV} 
\end{equation}

\ni and from Eq. (21) 

\begin{equation}
\mu^{(\nu)} \sim (6.5 - 6.6) \times 10^{-2}\;{\rm eV}\;.
\end{equation}

The neutrino mass formula (18) is not of the seesaw form. At any rate, no seesaw elements were used in its formulation. However, we constructed recently [4] a special seesaw neutrino model -- with the Dirac and Majorana neutrino mass matrices living in a specific "conspiracy"\, [5] -- that leads to the neutrino mass formula 

\begin{equation}
m_{\nu_i} = \mu^{(\nu)}\rho_i \left(1 + \frac{\varepsilon^{(\nu)} -1}{N^4_i} \right) \;,
\end{equation}

\ni where $\varepsilon^{(\nu)} > 0$ is a new constant. In Ref. [4] this constant gets the small value

\begin{equation}
\varepsilon^{(\nu)} \sim 7.35 \times 10^{-3} \;. 
\end{equation}

We can see that the mass spectra (18) and (31) are practically identical for $m_{\nu_2}$ and $m_{\nu_3}$, but differ for $m_{\nu_1}$ which becomes now nonzero since the mass formula (31) implies

\begin{equation}
m_{\nu_1} : m_{\nu_2} : m_{\nu_3} \simeq \varepsilon^{(\nu)} : 4 \!\cdot\! \frac{80}{81} : 24 \!\cdot\! \frac{624}{625} 
\end{equation}

\ni and

\begin{equation}
\mu^{(\nu)}\simeq \frac{29}{\varepsilon^{(\nu)} \!+\! 28}(m_{\nu_1} \!+\! m_{\nu_2} \!+\! m_{\nu_3}) \simeq \frac{29}{28}(m_{\nu_1} \!+\! m_{\nu_2} \!+\! m_{\nu_3}) = \frac{29}{\varepsilon^{(\nu)}}m_{\nu_1} = \frac{29}{4}\frac{81}{80}m_{\nu_2} = \frac{29}{24}\frac{625}{624} m_{\nu_3}\,.
\end{equation}

\ni From Eq. (33) 

\begin{equation}
\Delta m^2_{32}/\Delta m^2_{21} \simeq 36 \;,
\end{equation}

\ni what is larger by 1 than the value (22). Thus, using the experimental range (7) of $\Delta m^2_{21} $ and its experimental best fit (8) as an input, we obtain as a {\it prediction} the range very similar to (23):

\begin{equation}
2.6 <\Delta m^2_{32}/ (10^{-3}\;{\rm eV}^2) < 3.3 
\end{equation}

\ni and the best fit very similar to (24):

\begin{equation}
\Delta m^2_{32} \sim 2.9 \times 10^{-3}\;{\rm eV}^2 \;.
\end{equation}

\ni Notice that making use of the best fit (37) for $\Delta m^2_{32}$, we would predict from Eqs. (33) and (32)  

\begin{equation}
m_{\nu_1} \sim 1.7\times 10^{-5}\;{\rm eV}\;,\;m_{\nu_2} \sim 9.0 \times 10^{-3}\;{\rm eV} \;,\; m_{\nu_3} \sim 5.5\times 10^{-2}\;{\rm eV}
\end{equation}

\ni and from Eq. (34)

\begin{equation}
\mu^{(\nu)} \sim 6.6 \times 10^{-2}\;{\rm eV} \;.
\end{equation}

In conclusion, it is exciting that the generation-weighting factors $\rho_i$, so efficient in the case of charged-lepton masses, can be also useful for neutrino masses, namely, for predicting their ratio $\Delta m^2_{32}/\Delta m^2_{21}$ up to the deviation 35 -- 30 or 36 -- 30 from its actual experimental estimation 30 (its predicted value 35 or 36 still remains within the actual experimental limits $21< \Delta m^2_{32}/\Delta m^2_{21} < 42$). This suggests the hypothesis that the proposed simple mass formula (18) or its seesaw variation (31) describes, at least approximately, the true character of neutrino mass spectrum.

\vspace{0.2cm}

{\centerline{\bf Supplement}} 

\vspace{0.2cm}

One can achieve the full agreement with the actual experimental estimate $\Delta m^2_{32}/\Delta m^2_{21} \sim 30$ by introducing an appropriate second free parameter, but then the prediction for this ratio is lost. For example, the simplest mass formula

\begin{equation}
m_{\nu_i} = \mu^{(\nu)} \rho_i (1 - \beta \delta _{i3}) 
\end{equation}

\ni evolving from Eq. (9), where $\beta > 0$ is the second free parameter, gives

\begin{equation}
m_{\nu_1} : m_{\nu_2} : m_{\nu_3} = 1 : 4 : 24(1 - \beta)
\end{equation}

\ni and

\begin{equation}
\mu^{(\nu)} = \frac{29}{5+ 24(1-\beta)}(m_{\nu_1} + m_{\nu_2} + m_{\nu_3}) = 29m_{\nu_1} = \frac{29}{4} m_{\nu_2}  = \frac{29}{24(1-\beta)} m_{\nu_3} \,. 
\end{equation}

\ni From Eq. (41)

\begin{equation}
\Delta m^2_{32}/\Delta m^2_{21} =  \frac{16[36(1-\beta)^2 - 1]}{15} \;.
\end{equation}

\ni This leads to the value $\sim 30$ if

\begin{equation}
\beta \sim 0.10 \;.
\end{equation}

\ni So, $\beta $ is a small parameter ($\Delta m^2_{32}/\Delta m^2_{21} = 112/3 \simeq 37$ for $\beta = 0$).

Using the experimental range (7) of $\Delta m^2_{21}$ and its experimental best fit (8), one gets

\begin{equation}
2.2 < \Delta m^2_{32}/(10^{-3}\;\, {\rm eV}^2) < 2.7 
\end{equation}

\ni and

\begin{equation}
\Delta m^2_{32} \sim 2.4 \times 10^{-3}\;\, {\rm eV}^2 \;.
\end{equation}

\ni Then, from Eq. (41)

\begin{equation}
m_{\nu_1} \sim 2.3 \times 10^{-3}\;\, {\rm eV} \,,\, m_{\nu_2} \sim 9.2 \times 10^{-3}\;\, {\rm eV} \,,\, m_{\nu_3} \sim 5.0 \times 10^{-2}\;\, {\rm eV} 
\end{equation}

\ni and from Eq. (42)

\begin{equation}
\mu^{(\nu)} \sim 6.7 \times 10^{-2}\;\, {\rm eV} \;.
\end{equation}

\ni Here, the experimental estimates of $\Delta m^2_{21}$ and $\Delta m^2_{32}/\Delta m^2_{21}$ are both the input.

\vfill\eject

~~~~
\vspace{0.5cm}

{\centerline{\bf References}}

\vspace{0.5cm}

{\everypar={\hangindent=0.6truecm}
\parindent=0pt\frenchspacing

{\everypar={\hangindent=0.6truecm}
\parindent=0pt\frenchspacing

[1]~W. Kr\'{o}likowski, {\it Acta Phys. Pol.} {\bf B 33}, 2559 (2002) ({\tt hep--ph/0203107}); and references therein. 

\vspace{0.2cm}

[2]~Particle Data Group, {\it Review of Particle Physics, Phys.~Lett.} {\bf B 592} (2004).

\vspace{0.2cm}

[3]~For a review {\it cf. } J.W.F.~Valle, {\tt hep--ph/0410103}; D. Kie{\l}czewska, {\it Nucl. Phys.} {\bf B} ({\it Proc. Suppl.}) {\bf 136}, 77 (2004).

\vspace{0.2cm}

[4]~W. Kr\'{o}likowski, {\it Acta Phys. Pol.} {\bf B 35}, 673 (2004) ({\tt hep--ph/0401101}). 

\vspace{0.2cm}

[5]~S.M. Barr and I.~Dorsner, {\it Nucl. Phys.} {\bf B 585}, 79 (2000).

\vfill\eject

\end{document}